\documentclass[a4paper,11pt,nofootinbib,longbibliography]{article}
\usepackage{pos}

\usepackage{graphicx}
\usepackage{amsmath}
\usepackage{hyperref}

\title{Cosmological intercept tension}

\author[1,2]{Jia-Qi Wang}
\emailAdd{wangjiaqi@itp.ac.cn}
\author[*,1,3]{Shao-Jiang Wang}
\emailAdd{schwang@itp.ac.cn (corresponding author and speaker)}

\affiliation[1]{Institute of Theoretical Physics, Chinese Academy of Sciences (CAS), Beijing 100190, China}

\affiliation[2]{University of Chinese Academy of Sciences (UCAS), Beijing 100049, China}

\affiliation[3]{Asia Pacific Center for Theoretical Physics (APCTP), Pohang 37673, Korea}

\abstract{The long-standing tension in the Hubble constant $H_0$ has motivated extensive explorations of both new physics and observational systematics, for example, the late-time systematics in measuring the B-band absolute magnitude $M_B$ of type Ia supernovae, which is degenerated with $H_0$ via an intercept $-5a_B=M_B+5\lg (c/H_0/\mathrm{Mpc})+25$ in the linear relation $m_B=5\lg d_L(z)-5a_B$ between the apparent magnitude $m_B$ and logarithmic dimensionless luminosity distance $\lg d_L(z)$. Therefore, this intercept can be evaluated directly from pure observational quantities ($m_B$ and the redshift $z$) for a given model of $d_L(z)$ without knowing underlying systematics in $M_B$-$H_0$ degeneracy. Hence, the constancy of this intercept across different supernova datasets and different redshift bins within the same dataset for a given late-time model serves as a powerful diagnostic for disentangling late-time new physics from local supernova systematics. In this mini-review, we will show that: (1) there is a local $a_B$ tension in PantheonPlus around $z\sim0.01$, and the elimination of it leads to a measurement $H_0=73.4\pm1.0\;\mathrm{km/s/Mpc}$, consistent with both SH0ES typical three-rung and first two-rung measurements; (2) there is a late-time $a_B$ tension in DES-Y5 around $z\sim0.1$, and the elimination of it largely reduces the preference for dynamical dark energy. We also update the late-time $a_B$-tension analysis for both DES-Y5 and DES-Dovekie supernovae, and find that this $a_B$ tension around $z\sim0.1$ is mainly driven by the inter-data tension between DES supernovae and DESI+Planck constraint, and the dynamical dark energy is preferred as a compromise of this tension. Finally, we briefly mention an interacting dark energy model that resolves this tension among DES, DESI, and Planck, and point out a crucial difference between the effective and apparent equations of state of dark energy when interpreting the data with a specific parameterization.}

\FullConference{Proceedings of the Corfu Summer Institute 2025, \href{https://indico.cern.ch/event/1544691/overview}{Tension in Cosmology 2025}, Mon-Repos, Corfu, Greece. Based on the \href{https://indico.cern.ch/event/1544691/contributions/6668222/}{invited talk} ``Cosmological tension implications for a non-minimally coupled dark matter with dark energy'' but slightly updated with DES-Dovekie.}

\tableofcontents

\begin{document}
\maketitle

\section{Introduction}

The standard $\Lambda$CDM model~\cite{Planck:2013pxb} has achieved remarkable success in describing the observed Universe across a wide range of scales and epochs~\cite{Moresco:2022phi}. Nevertheless, several persistent discrepancies have raised the possibility that this concordance framework is incomplete~\cite{Perivolaropoulos:2021jda}. Among them, the Hubble tension~\cite{Bernal:2016gxb,Verde:2019ivm,Knox:2019rjx,Riess:2020sih,Freedman:2021ahq} remains the most prominent: local distance-ladder measurements calibrated by SH0ES~\cite{Riess:2021jrx} favor a significantly higher value of the Hubble constant $H_0$ than that inferred from Planck CMB data under $\Lambda$CDM~\cite{Planck:2018vyg}. This discrepancy~\cite{DiValentino:2020zio,DiValentino:2021izs,Schoneberg:2021qvd,Shah:2021onj,Abdalla:2022yfr,Hu:2023jqc,Vagnozzi:2023nrq,Cai:2023sli,CosmoVerseNetwork:2025alb} has motivated a large literature on early-time new physics~\cite{Krishnan:2020obg,Jedamzik:2020zmd,Hill:2020osr,Lin:2021sfs,Vagnozzi:2021gjh,Philcox:2022sgj}, late-time new physics~\cite{Benevento:2020fev,Camarena:2021jlr,Efstathiou:2021ocp,Cai:2021wgv,Cai:2021weh,Cai:2022dkh,Calza:2025yfm}, and unresolved systematics~\cite{Yu:2022wvg,Huang:2024erq,Huang:2024gfw,Huang:2025som}.

A useful way to sharpen this discussion is to distinguish different projections of the Hubble tension. If one attributes the discrepancy to early-Universe physics while keeping the standard late-Universe, the problem appears as a sound-horizon $r_s$ tension~\cite{Bernal:2016gxb,Verde:2019ivm,Knox:2019rjx,Riess:2020sih}. The reduction of $r_s$ tension is achieved at a price of pushing the scalar spectral index closer to the Harrison-Zeldovich scale-invariant spectrum via $\delta n_s\sim0.4\delta H_0/H_0$~\cite{Ye:2021nej} (see, e.g., Refs.~\cite{Fu:2023tfo,Fu:2025ciy} for a single-field inflationary reconcilement of Harrison-Zeldovich spectrum), and also enlarging the matter fluctuation $S_8$ tension, as either speeding up the early-Universe expansion history or advancing the recombination history would all require a larger matter fraction $\Omega_m$ (thus a larger $S_8$) to match either galaxy clustering or lensing data~\cite{Jedamzik:2020zmd}. This observation was originally recognized as some sort of ``early-time No-Go theorems''~\cite{Vagnozzi:2023nrq}, but intriguingly, the recent ACT preference for a larger $n_s$~\cite{AtacamaCosmologyTelescope:2025blo,AtacamaCosmologyTelescope:2025nti} and recent KiDs-Legacy rescue of $S_8$ tension~\cite{Wright:2025xka} seem to reassess the early-Universe solutions~\cite{Poulin:2025nfb,SPT-3G:2025vyw,Toda:2025kcq,Wang:2025djw,Yin:2026gss}.

If instead one assumes the early Universe is standard but seeks a late-time explanation, similar ``late-time No-Go theorems''~\cite{Benevento:2020fev,Camarena:2021jlr,Efstathiou:2021ocp,Cai:2021weh,Cai:2022dkh,Huang:2024erq} arise as results of strong constraints from inverse-distance-ladder measurements~\cite{Cuesta:2014asa,Heavens:2014rja,Aubourg:2014yra,Verde:2016ccp,Alam:2016hwk,Verde:2016wmz,Macaulay:2018fxi,Feeney:2018mkj,Lemos:2018smw,eBOSS:2020yzd,Ling:2025lmw}, which appear as an absolute-magnitude $M_B$ tension. This $M_B$ tension is found to be degenerated with $H_0$ tension through an intercept $a_B$ tension~\cite{Huang:2024erq,Huang:2024gfw,Huang:2025som,Ling:2025lmw} when fitting the linear relation between the apparent magnitude $m_B$ and the logarithmic distance $\lg d_L(z)$,
\begin{align}
m_B = 5\lg d_L(z)-5a_B, \qquad
-5a_B = M_B + 5\lg \left(\frac{c/H_0}{\mathrm{Mpc}}\right)+25,
\end{align}
where $m_B$ is the corrected apparent magnitude after standardization and $d_L(z)\equiv D_L(z)/(c/H_0)$ is the dimensionless luminosity distance at a reshift $z$ for a given late-time model $D_L(z)=c(1+z)\int_0^z\mathrm{d}z'/H(z')$. Since $a_B$ can be directly reconstructed from supernova Hubble-diagram data, it separates the directly observed supernova Hubble diagram from the model-dependent calibration in $H_0$ and $M_B$, and hence offers a clean consistency test between data and cosmological modeling.

In this proceeding review, we use the intercept $a_B$ diagnostics as a unifying probe to constrain our late and local Universe. Our main message is that $a_B$ is not merely a degeneracy direction between $H_0$ and $M_B$, but a convenient and powerful diagnostic that distinguishes late-time new physics from low-redshift supernova systematics. We first review the local $a_B$ tension~\cite{Huang:2024erq} in PantheonPlus~\cite{Riess:2021jrx,Brout:2022vxf,Scolnic:2021amr,Brout:2021mpj,Peterson:2021hel,Carr:2021lcj,Popovic:2021yuo} and explain how its elimination narrows the Hubble tension to the first two rungs of the distance ladder~\cite{Huang:2024gfw}. We then turn to the late-time $a_B$ tension~\cite{Huang:2025som,Ling:2025lmw} in DES-Y5~\cite{DES:2024jxu,DES:2024hip,DES:2025tir} and its updated comparison with DES-Dovekie~\cite{DES:2025sig}, showing that the apparent preference for dynamical dark energy can largely be understood as a compromise fit to the inter-data tension among DES supernovae, DESI BAO~\cite{DESI:2024mwx,DESI:2024hhd}, and Planck CMB. Finally, we briefly comment on the possibility that an interacting dark-energy scenario~\cite{Wang:2025znm} may provide a more suitable framework for reconciling the remaining dataset inconsistencies and reproducing the apparent crossing behavior of the dark-energy equation of state.

\section{Local $a_B$ tension}\label{sec:local}

The studies of the intercept ($a_B$) tension originated from formulating our previous late-time ``No-Go theorem''~\cite{Cai:2021weh,Cai:2022dkh} in a fully model-independent manner~\cite{Huang:2024erq}, that is, fitting a fully model-independent Parameterization based on the cosmic Age (PAge~\cite{Huang:2020mub,Luo:2020ufj,Huang:2020evj,Huang:2021aku,Huang:2021tvo,Huang:2022txw,Li:2022inq,Wang:2023mir,Wang:2024nsi,Wang:2024rus,Yao:2024kex}) to a fully model-independent inverse-distance-ladder (IDL=HFSN+2DBAO+CC) consisting of cosmologically model-independent Hubble-flow supernovae (HFSN) and two-dimensional baryon acoustic oscillation (2DBAO) calibrated by cosmologically model-independent high-redshift calibrators from cosmic chronometer (CC). Here, the PAge model serves as a global, economic, and precise parameterization in the late/local Universe for any homogeneous modification gently beyond the $\Lambda$CDM model, while for those more abrupt changes in the expansion rate that cannot be well captured by the PAge model, a specific phantom-dark-energy (PDE) model following Ref.~\cite{Efstathiou:2021ocp} is chosen as a representative illustration in Ref.~\cite{Huang:2024erq}. The first half of Ref.~\cite{Huang:2024erq} strengthens the previous ``No-Go theorem'' for the late Universe with strong evidence ($\Delta\mathrm{BIC}=6.27$) against the PAge model and decisive evidence ($\Delta\mathrm{BIC}=18.42$) against the PDE model over the $\Lambda$CDM model. 

Intriguingly, when calibrating the local-Universe SNe with the above IDL constraints from the late Universe, the intercept ($a_B$) tension arises as we will below.

\subsection{Local vs. Hubble-flow supernovae}\label{subsec:locallate}

\begin{figure}
    \centering
    \includegraphics[width=0.53\linewidth]{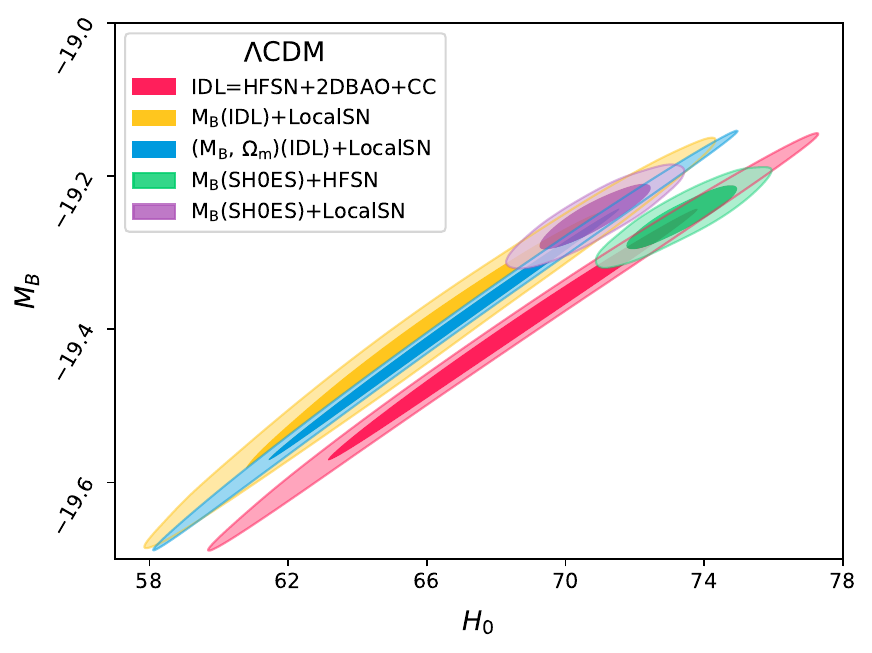}
    \includegraphics[width=0.46\linewidth]{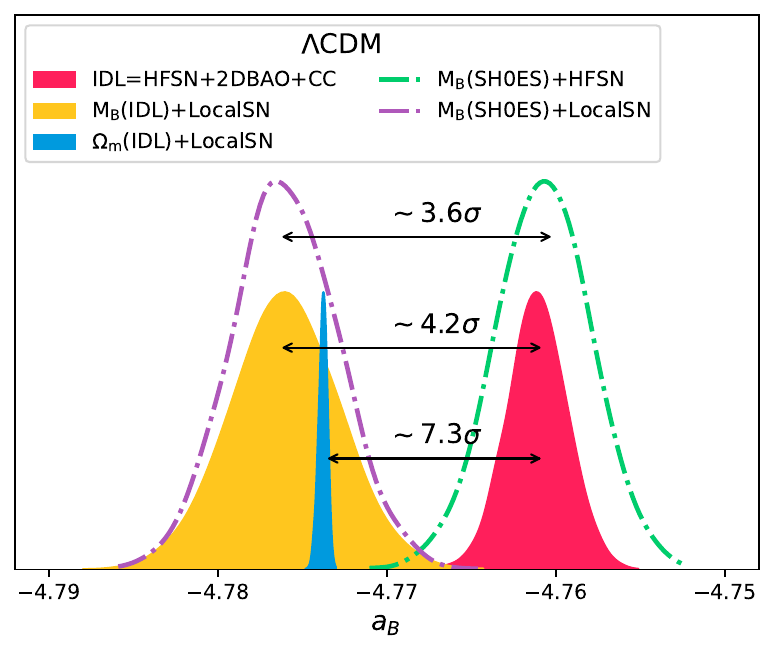}\\
    \includegraphics[width=1\linewidth]{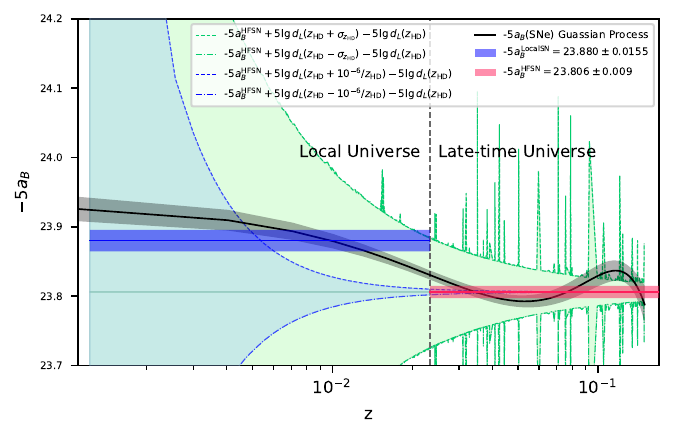}
    \caption{Local versus Hubble-flow supernova intercept constraints from Ref.~\cite{Huang:2024erq}. The upper panels illustrate the displacement between the local-SN and Hubble-flow/IDL constraints in the $H_0$--$M_B$ plane and in the corresponding $a_B$ posterior. Although $H_0$ and $M_B$ are individually consistent within uncertainties, the contours are separated along the diagonal intercept direction. The lower panel visualizes the same discrepancy directly in $-5a_B$, showing that the local and Hubble-flow supernovae prefer different intercepts of the magnitude--redshift relation. The volume effect of SN samples and the peculiar-velocity effect in the model are shown in light blue- and green-shaded regions with $1\sigma$ errors, respectively. The light black-shaded region is directly reconstructed from the SN data using the Gaussian process. }
    \label{fig:2.1}
\end{figure}

Let us first compare SNe Ia inside and outside the local inhomogeneous region. The PantheonPlus sample is split into a local part and a Hubble-flow part within the redshift range $0.0233<z<0.15$, where the lower bound roughly corresponds to the homogeneity scale, $R_{\rm homo}\simeq 70\,{\rm Mpc}/h$, and suppresses the impact of local cosmic variance and peculiar velocities, while the upper bound keeps the sample sufficiently close to the linear Hubble diagram so that it is insensitive to detailed late-time dark-energy evolution. This gives 490 Hubble-flow SNe Ia from PantheonPlus. The complementary local sample contains 336 SNe Ia below the homogeneity scale, $z<0.0233$, with peculiar-velocity corrections treated separately in the PantheonPlus covariance matrix~\cite{Riess:2021jrx,Brout:2022vxf,Scolnic:2021amr,Peterson:2021hel,Carr:2021lcj}.

The Hubble-flow (HF) sample was combined with cosmologically model-independent 2DBAO and CC data to form an IDL (=HFSN+2DBAO+CC) constraint, For $\Lambda$CDM, this gives
\begin{align}
    H_0 = 68.5\pm 3.5~{\rm km\,s^{-1}\,Mpc^{-1}},
    \qquad
    M_B=-19.40\pm0.11,
\end{align}
with very similar constraints for the PAge and PDE parameterizations. Calibrating the local SNe with the IDL posterior on $M_B$ instead gives, for $\Lambda$CDM,
\begin{align}
    H_0 = 66.2\pm3.4~{\rm km\,s^{-1}\,Mpc^{-1}},
    \qquad
    M_B=-19.40\pm0.11 .
\end{align}
Therefore, neither $H_0$ nor $M_B$ alone displays a significant local-late discrepancy in this calibration scheme. The important point is instead that the two posterior contours are displaced almost along the diagonal degeneracy direction in the $H_0$--$M_B$ plane, as shown in the first panel of Fig.~\ref{fig:2.1}. This diagonal direction is precisely the supernova intercept $a_B$.

The pure IDL constraint gives, for example,
\begin{align}
    a_B^{\rm IDL}=-4.7612\pm0.0018
    \qquad
    (\Lambda{\rm CDM}),
\end{align}
whereas the IDL-calibrated local supernovae give
\begin{align}
    a_B^{\rm local}=-4.7761\pm0.0031
\end{align}
if one first uses the IDL $M_B$ prior to infer $H_0$ from the local SNe. Equivalently, one may reconstruct $a_B$ directly from the local Hubble diagram through~\cite{Efstathiou:2021ocp}
\begin{align}
    a_B=
    \frac{
    \sum_{ij}C^{-1}_{ij}
    \left[
        \lg d_L(z_i;\{p_\alpha\})
        -\frac15 m_{B,i}
    \right]
    }{
    \sum_{ij}C^{-1}_{ij}
    },
    \label{eq:local-aB-direct}
\end{align}
where $C_{ij}$ is the PantheonPlus covariance matrix and the cosmological parameters $\{p_\alpha\}$ are calibrated by the IDL posterior. This direct reconstruction gives
\begin{align}
    a_B^{\rm local,direct}=-4.7739^{+0.00027}_{-0.00023}
    \qquad
    (\Lambda{\rm CDM}),
\end{align}
again clearly separated from the Hubble-flow/IDL value. Similar separation in the supernova intercept also manifests for both PAge and PDE models, indicating a model-insensitive trend in this intercept $a_B$ tension. 

Thus, the local and Hubble-flow SNe are not primarily in tension in $H_0$ or $M_B$ separately; rather, they are in tension in the directly observable intercept $a_B$ of the magnitude--redshift relation. For $\Lambda$CDM and PAge, the discrepancy is at the level of $3\sim7\sigma$, depending on whether the local inference is implemented indirectly through the IDL $M_B$ prior or directly through Eq.~\eqref{eq:local-aB-direct}. This diagonal separation appears independent of calibrations, as it also occurs between the SH0ES-calibrated local SN and HF SN samples, as shown in the first panel of Fig.~\ref{fig:2.1}. In terms of the plotted intercept $-5a_B$ in the second panel of Fig.~\ref{fig:2.1}, the shift is about $0.06$ mag, which is small in absolute magnitude but highly significant compared with the statistical precision of the local Hubble diagram. In short, this local $a_B$ tension isolates a model-insensitive, calibration-independent inconsistency in the supernova magnitude--redshift relation, with an intercept that absorbs the usual $H_0$--$M_B$ degeneracy. 

Once a direct discrepancy in $a_B$ is identified across $\Lambda$CDM, PAge, and PDE, and under both IDL and SH0ES calibrations, its origin can no longer be attributed simply to the absolute-magnitude calibration $M_B$. Instead, it must arise either from new physics or systematics affecting the luminosity-distance relation $d_L(z)$, or from uncertainties in the measured redshifts $z_i$, especially peculiar-velocity effects at very low redshift $z\lesssim0.01$. The third panel of Fig.~\ref{fig:2.1} illustrates that ordinary volumetric or Malmquist selection effects are far too small to explain the local--late intercept mismatch, while the estimated redshift uncertainties from peculiar velocities are already included in the covariance matrix. However, a hypothetical effect that mimics an unmodeled peculiar-velocity-like redshift shift could, in principle, cover the observed $a_B$ tension. A Gaussian-process reconstruction of the intercept evolution from Pantheon+ further supports a data-driven discrepancy between the local and late-Universe supernova intercepts.

\subsection{First two rungs vs. third-rung ladders}\label{subsec:firsttwothird}

\begin{figure}
    \centering
    \includegraphics[width=0.56\linewidth]{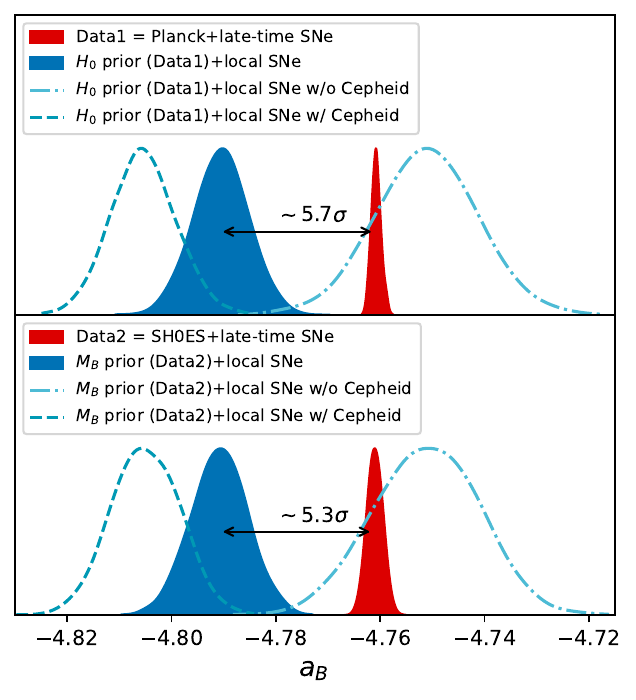}
    \includegraphics[width=0.43\linewidth]{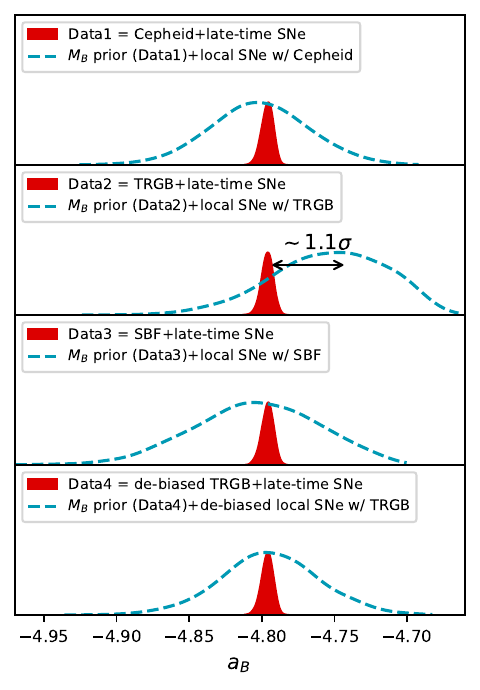}\\
    \includegraphics[width=1\linewidth]{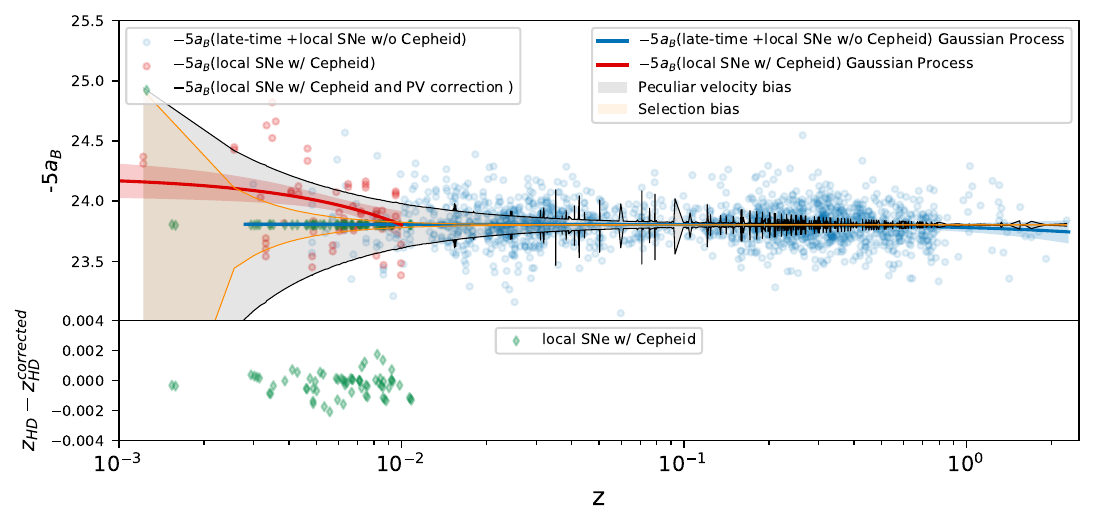}
    \caption{Second-rung versus third-rung consistency test from Ref.~\cite{Huang:2024gfw}. The first panel shows that the Cepheid-hosted local SNe, which calibrate $M_B$ in the second rung, are responsible for the dominant local displacement in $a_B$, while the non-Cepheid local SNe remain consistent with the third-rung Hubble-flow intercept. The second panel shows the $a_B$ Gaussian posteriors for late-time and local CSP SNe Ia calibrated by Cepheid/TRGB/SBF/de-biased TRGB. The bottom panel shows the intercept $-5a_B$ for the whole PantheonPlus SNe sample. Blue points denote third-rung SNe over $z\in[0.003,2.3]$, while red points denote second-rung Cepheid-hosted SNe. The blue and red bands show the corresponding Gaussian-process reconstructions; gray and orange bands indicate the expected peculiar-velocity and selection biases~\cite{Huang:2024erq}. Red diamonds mark Cepheid-hosted SNe after $a_B$-consistency PV corrections for the redshifts as indicated by green diamonds.}
    \label{fig:2.2}
\end{figure}

The above local $a_B$ tension can be sharpened further by asking which part of the distance ladder is responsible for the intercept mismatch. In the conventional SH0ES three-rung ladder~\cite{Riess:2021jrx}, the first rung calibrates Cepheid distances, the second rung uses Cepheid-hosted SNe Ia to infer the standardized SN absolute magnitude $M_B$, and the third rung uses Hubble-flow SNe Ia to determine the intercept $a_B$. The final inference of $H_0$ combines these two ingredients through
\begin{align}
    -5a_B^{\rm 3rd}=M_B^{\rm 2nd}+5\lg\left(\frac{c/H_0}{\rm Mpc}\right)+25 .
\end{align}
This procedure is internally consistent only if the Cepheid-hosted second-rung SNe and the third-rung SNe share the same standardized magnitude--redshift relation, or equivalently the same intercept $a_B$ after all local redshift and magnitude corrections have been applied.

To locate the origin of the mismatch, Ref.~\cite{Huang:2024gfw} divided the low-redshift PantheonPlus SNe into Cepheid-hosted local SNe and non-Cepheid-hosted local SNe. The non-Cepheid-hosted local SNe, which belong to the same SN population as the third-rung Hubble-flow sample, are consistent with the late-time third-rung intercept. By contrast, the Cepheid-hosted second-rung SNe show the dominant displacement in $a_B$, as illustrated in the first panel of Fig.~\ref{fig:2.2}. Therefore, the local $a_B$ tension is more specifically a tension between the Cepheid-hosted second-rung SNe and the third-rung SNe, rather than a generic discrepancy between all local and Hubble-flow SNe.

One possible interpretation would be an unaccounted apparent-magnitude systematic, or equivalently an effective transition in $M_B$ between the local and Hubble-flow samples. However, this explanation is disfavored by directly comparing the Cepheid distance moduli with the SN apparent magnitudes. For the most discrepant local Cepheid-hosted subsample with $z_{\rm HD}<0.007$, one finds
\begin{align}
    M_B = -19.235\pm0.035 ,
\end{align}
which is consistent with the global Cepheid calibration
\begin{align}
    M_B = -19.251\pm0.031 .
\end{align}
Thus, the observed $a_B$ displacement does not correspond to a genuine brightening of the Cepheid-calibrated SNe. The more plausible origin is instead a redshift systematic, especially the peculiar-velocity correction in the very nearby Universe, as shown in the bottom panel of Fig.~\ref{fig:2.2}.

A simple way to restore the second--third-rung consistency is to impose the observed stability of $a_B$ as a redshift-correction condition. For each local Cepheid-hosted SN, one scans the allowed redshift interval $z_i\in\left[z_{\rm HD}-3\sigma_{z_{\rm HD}},z_{\rm HD}+3\sigma_{z_{\rm HD}}\right]$, and computes $a_{B,i}=\lg d_L(z_i)-0.2\,m_{B,i}$. The corrected redshift $z_{\rm HD}^{\rm corrected}$ is then chosen as the value whose $a_{B,i}$ is closest to the late-time third-rung intercept. This procedure effectively uses the more reliable third-rung stability of the Hubble diagram to correct the peculiar-velocity-induced redshift bias of the less reliable second-rung SNe.
Once this $a_B$ consistency is imposed, the Cepheid distance moduli alone can be used to infer $H_0$ without using the Hubble-flow SNe as a third rung. The likelihood is $\ln\mathcal{L}=-\frac12\Delta\mu^TC_{\rm stat.+syst.}^{-1}\Delta\mu$ with $\Delta\mu_i=\mu_{{\rm Cepheid},i}-\mu_{\rm model}\left(z_{{\rm HD},i}^{\rm corrected};H_0\right)$. This first-two-rung inference gives
\begin{align}
    H_0=73.4\pm1.0\;{\rm km\,s^{-1}\,Mpc^{-1}},
\end{align}
in agreement with both the usual SH0ES three-rung result~\cite{Riess:2021jrx} and the independent two-rung analysis of Ref.~\cite{Kenworthy:2022jdh} with rigorous PV correction procedure. Without the redshift correction, the same exercise would give an anomalously low value, $H_0=66.36\pm0.92\;{\rm km\,s^{-1}\,Mpc^{-1}}$, which signals the importance of peculiar-velocity-induced redshift systematics for the nearby Cepheid-hosted sample.

This conclusion was further checked with the Carnegie Supernova Project (CSP) sample, where an $a_B$ discrepancy is only exhibited for TRGB calibration as shown in the second panel of Fig.~\ref{fig:2.2}. Using local CSP SNe hosted by Cepheid, TRGB, and SBF calibrators, and then fitting directly to the calibrator distance moduli and local host redshifts without invoking third-rung SNe, one obtains
\begin{align}
H_0
=
\begin{cases}
73.1\pm2.4\;{\rm km\,s^{-1}\,Mpc^{-1}}, 
& z_{\rm CSP}+\mu_{\rm Cepheid},\\
74.5\pm3.5\;{\rm km\,s^{-1}\,Mpc^{-1}}, 
& z_{\rm CSP}+\mu_{\rm TRGB},\\
72.1\pm2.3\;{\rm km\,s^{-1}\,Mpc^{-1}}, 
& z_{\rm CSP}+\mu_{\rm SBF}.
\end{cases}
\end{align}
In particular, our restoration of $a_B$ consistency essentially improves Ref.~\cite{Freedman:2024eph} to match Ref.~\cite{Riess:2024vfa} for the TRGB case. All three calibrators prefer a high local value of $H_0$, independently of the third-rung Hubble-flow SNe.  Therefore, the high SH0ES-like value of $H_0$ is unlikely to originate from a late-time third-rung SN systematic such as an unmodeled transition in $M_B$. Instead, the Hubble tension is narrowed down to a discrepancy between Planck-CMB constraints and the first two rungs of the local distance ladder.

\section{Late-time $a_B$ tension}\label{sec:late}

Having found and corrected the $a_B$ tension in the local Universe around $z\sim0.01$, we then move to test the $a_B$ tension in the late Universe around $z\sim0.1$.

\subsection{PantheonPlus vs. DES-Y5 supernovae}\label{subsec:PPDESY5}

\begin{figure}
    \centering
    \includegraphics[width=1\linewidth]{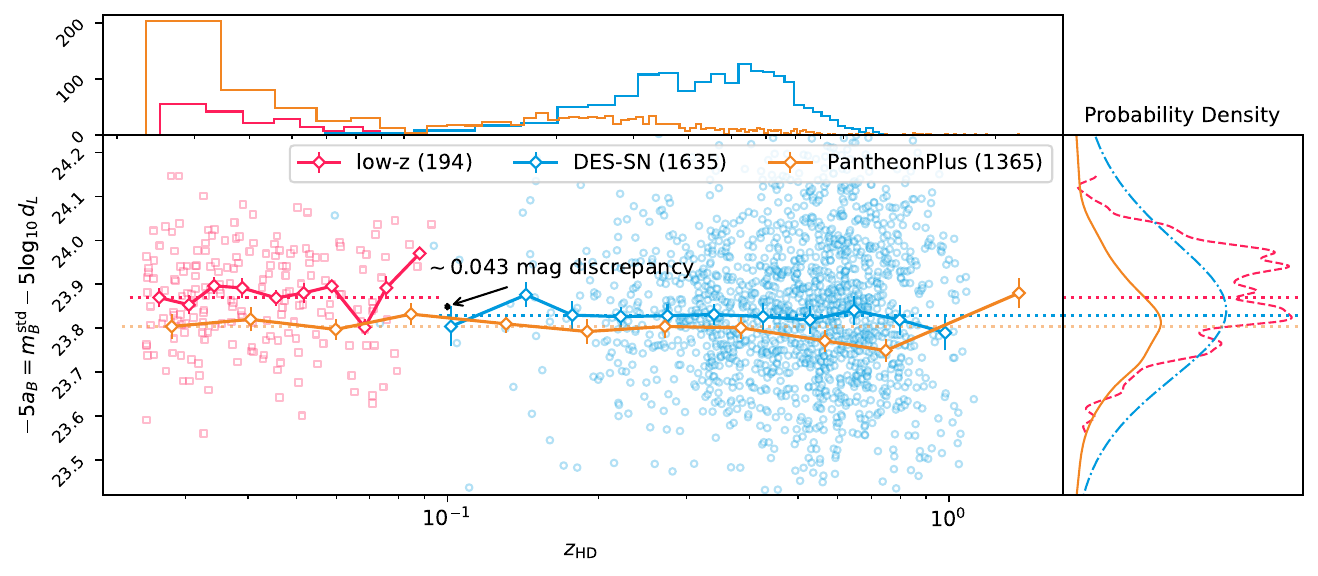}\\
    \includegraphics[width=1\linewidth]{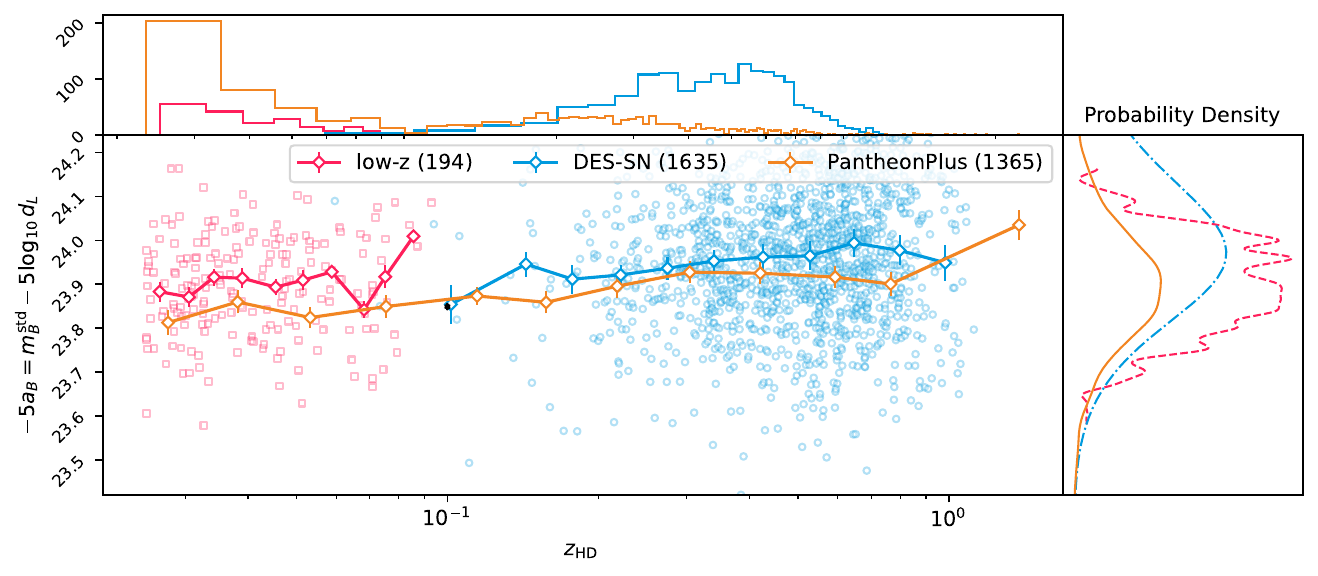}\\
    \includegraphics[width=0.49\linewidth]{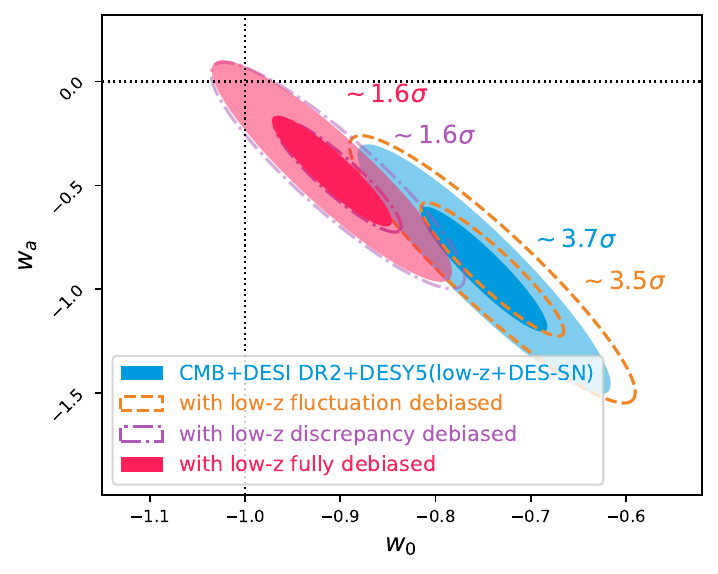}
    \includegraphics[width=0.49\linewidth]{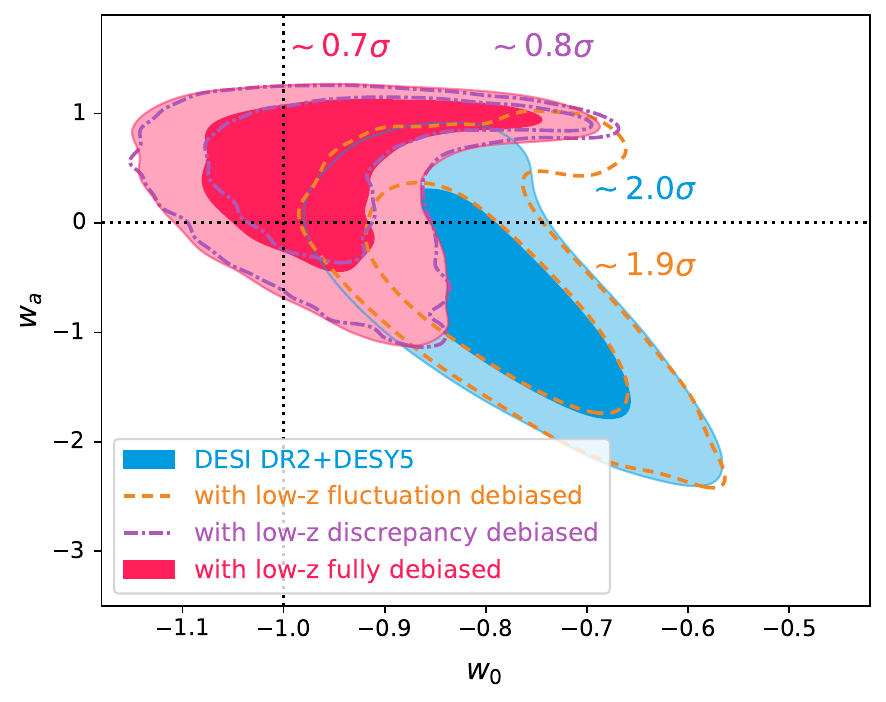}\\
    \caption{
    The late-time $a_B$ diagnosis for DES-Y5 compared with PantheonPlus from Ref.~\cite{Huang:2025som}. In the first panel, the DES-SN subset has a stable intercept compatible with a single magnitude--distance relation just like the PantheonPlus sample, while the external low-$z$ subset shows large fluctuations and a mean offset of about $0.043$ mag. relative to DES-SN. This cannot be cured simply by turning into another model without spoiling the PantheonPlus stability in $a_B$, as shown in the second panel for a best-fit $w_0w_a$CDM model  from DESI DR2~\cite{DESI:2025zgx} with $\Omega_m=0.352_{-0.018}^{+0.041}$, $w_0=-0.48_{-0.17}^{+0.35}$, and $w_a<-1.34$. In the third panel, restoring the low-$z$/DES-SN $a_B$ consistency substantially reduces the apparent preference for CPL-like dynamical dark energy in Planck+DESI+DES-Y5.
    }
    \label{fig:3.1}
\end{figure}

We now turn from the local $a_B$ tension in the Hubble-constant problem to the late-time $a_B$ tension~\cite{Huang:2025som} relevant for the recent DESI preference for dynamical dark energy. The DESI DR1/DR2 result~\cite{DESI:2024mwx,DESI:2024hhd}, when combined with Planck CMB and DES-Y5 SNe, gives an apparent preference for a Chevallier-Polarski-Linder (CPL~\cite{Chevallier:2000qy,Linder:2002et})-like dynamical dark-energy equation of state with phantom crossing. However, this statement depends sensitively on which SN compilation is used. Replacing PantheonPlus by DES-Y5 increases the significance of the deviation from $\Lambda$CDM, while removing the external low-$z$ SN subset from DES-Y5 largely weakens this preference. This immediately suggests that the key issue is not simply DESI BAO alone, but the consistency between DESI+Planck and the low-redshift anchoring part of DES-Y5.

The DES-Y5 compilation consists of the photometrically classified DES-SN sample together with an external low-$z$ SN sample. The latter contains 194 SNe Ia, including CSP, CfA, and Foundation objects, and is therefore more heterogeneous than the DES-SN part~\cite{DES:2024jxu,DES:2024hip}. By contrast, PantheonPlus contains many more low-redshift SNe and was constructed from a broader but carefully cross-calibrated set of subsamples~\cite{Riess:2021jrx,Brout:2022vxf,Scolnic:2021amr,Brout:2021mpj,Peterson:2021hel,Carr:2021lcj,Popovic:2021yuo}. Since many DES-Y5 low-$z$ SNe overlap with PantheonPlus, the difference between the two compilations is not merely one of sky coverage or redshift range, but of calibration, standardization, and covariance treatment.

Similar to PantheonPlus SNe analysis, for DES-Y5, we use the standardized apparent magnitudes after stretch, color, selection-bias, light-curve-fitting, and host-mass corrections, and write the distance-modulus residual vector as $\Delta \vec D = m_B^{\rm std}-M_B-\mu_{\rm model}$ with likelihood $-2\ln{\cal L} = \Delta \vec D^{\,T} C_{\rm stat+syst}^{-1} \Delta \vec D$.
The same intercept diagnostic used in the local $a_B$ tension analysis can then be applied to DES-Y5 from the magnitude-redshift relation $m_{B,i}^{\rm std} = 5\lg d_L(z_i)-5a_B$ with an intercept $-5a_B = M_B+5\lg\left(\frac{c/H_0}{\rm Mpc}\right)+25$. For a fixed fiducial late-time cosmology, any redshift-dependent inconsistency in $-5a_{B,i}=m_{B,i}^{\rm std}-5\lg \hat d_L(z_i)$ signals either unaccounted SN systematics or a genuine failure of the assumed homogeneous expansion history. 

The comparison is sharp as shown in the first panel of Fig.~\ref{fig:3.1}. PantheonPlus exhibits an approximately globally stable intercept across its low- and high-redshift parts. The DES-SN component of DES-Y5 also shows a stable $a_B$ distribution. However, the external low-$z$ subset of DES-Y5 shows both large irregular fluctuations and a weighted-average offset of about $0.043$ mag. relative to the DES-SN part. This offset is close to the magnitude residual difference previously noticed in the PantheonPlus--DES-Y5 comparison~\cite{Efstathiou:2024xcq}, although our interpretation is slightly different: the relevant quantity is not only the offset between two different compilations, but the internal low-$z$/high-$z$ inconsistency within DES-Y5 itself. Even after the DES-Y5 reanalysis in Ref.~\cite{DES:2025tir}, which reduces part of the PantheonPlus--DES-Y5 offset by improving intrinsic-scatter and selection-function modeling, a residual low-$z$/high-$z$ mismatch inside DES-Y5 remains relevant for the cosmological inference.

This point is important for distinguishing new physics from systematics. The DES-Y5 low-$z$ intercept shift appears around $z\sim0.1$, well above the very local inhomogeneous regime but still in a range where PantheonPlus has sufficient statistics. If the DES-Y5 feature were caused by a genuine homogeneous transition in the luminosity distance, the same transition should also distort PantheonPlus. Instead, PantheonPlus remains internally consistent. Therefore, a homogeneous late-time transition that repairs DES-Y5 would generally spoil PantheonPlus, as shown in the second panel of Fig.~\ref{fig:3.1}, for the best-fit $w_0w_a$CDM model from DESI DR2~\cite{DESI:2025zgx} with $\Omega_m=0.352_{-0.018}^{+0.041}$, $w_0=-0.48_{-0.17}^{+0.35}$, and $w_a<-1.34$. This is the central reason why the DES-Y5 $a_B$ anomaly is more naturally interpreted as a low-$z$ SN-systematics issue than as genuine evidence for dynamical dark energy.

This conclusion is also visible directly in the cosmological fits in Ref.~\cite{Huang:2025som}. With Planck CMB, eBOSS BAO, and PantheonPlus, the preference for dynamical dark energy is only about $1.7\sigma$. Replacing eBOSS by DESI DR1 raises it to about $2.5\sigma$, and replacing PantheonPlus by DES-Y5 raises it to about $2.9\sigma$. When both replacements are made, Planck+DESI DR1+DES-Y5 reaches about $3.5\sigma$. However, once the external low-$z$ DES-Y5 SNe are removed, the significance drops back to about $1.7\sigma$. Similarly, Planck+DES-SN alone is consistent with $\Lambda$CDM at the $\sim0.6\sigma$ level, whereas Planck+low-$z$ SNe alone already prefers a direction with $w_0>-1$ and $w_a<0$. Hence, the apparent DES-Y5 contribution to dynamical dark energy is largely driven by the external low-$z$ SN anchor rather than by the homogeneous DES-SN sample itself.

A simple phenomenological correction is therefore to restore $a_B$ consistency for the external low-$z$ SNe from the more reliable DES-SN. In practice, one may smooth the bin-to-bin low-$z$ fluctuations,
\begin{align}
m_{B,i\in{\rm low}-z}^{\rm fluc-debias} = 5\lg d_L(z_i) - 5\overline{a_{B,{\rm low}-z}},
\end{align}
or correct the mean low-$z$/DES-SN discrepancy by shifting the low-$z$ standardized magnitudes by the observed intercept mismatch,
\begin{align}
m_{B,i\in{\rm low}-z}^{\rm disp-debias} = m_{B,i\in{\rm low}-z}^{\rm std} - 0.043 .
\end{align}
Combining both gives the fully debiased low-$z$ sample,
\begin{align}
m_{B,i\in{\rm low}-z}^{\rm full-debias} = 5\lg d_L(z_i) - 5\overline{a_{B,{\rm DES-SN}}}.
\end{align}
As shown in the third panel of Fig.~\ref{fig:3.1}, the mean-offset correction alone is already sufficient to remove most of the dynamical-dark-energy preference. For Planck+DESI DR1/DR2+DES-Y5, the original $\sim3.5\sigma/\sim3.7\sigma$ preference is reduced to only $\sim1.5\sigma/\sim1.6\sigma$ after restoring the low-$z$/DES-SN intercept consistency. Without Planck, the corresponding late-Universe-only preference is reduced even further to below the $1\sigma$ level. The later DESI DR2 analysis also provided a related check: removing DES-Y5 SNe below $z\simeq0.1$ reduces the dynamical-dark-energy preference to about $2\sigma$~\cite{DESI:2025zgx}.

Therefore, the comparison between PantheonPlus and DES-Y5 teaches a lesson parallel to that from the local $a_B$ tension. The intercept $a_B$ is not merely a nuisance degeneracy between $H_0$ and $M_B$; it is a sensitive internal-consistency diagnostic of SN compilations. PantheonPlus and the DES-SN subset pass this diagnostic rather well, whereas the external low-$z$ DES-Y5 subset does not. Consequently, the DESI+Planck+DES-Y5 preference for dynamical dark energy should be interpreted cautiously: before invoking a late-time transition in the cosmic expansion history, one must first account for the late-time $a_B$ mismatch introduced by the low-redshift DES-Y5 anchor.

\subsection{PantheonPlus vs. DES-Y5/DES-Dovekie}\label{subsec:PPDESY5D}

\begin{figure}
    \centering
    \includegraphics[width=1\linewidth]{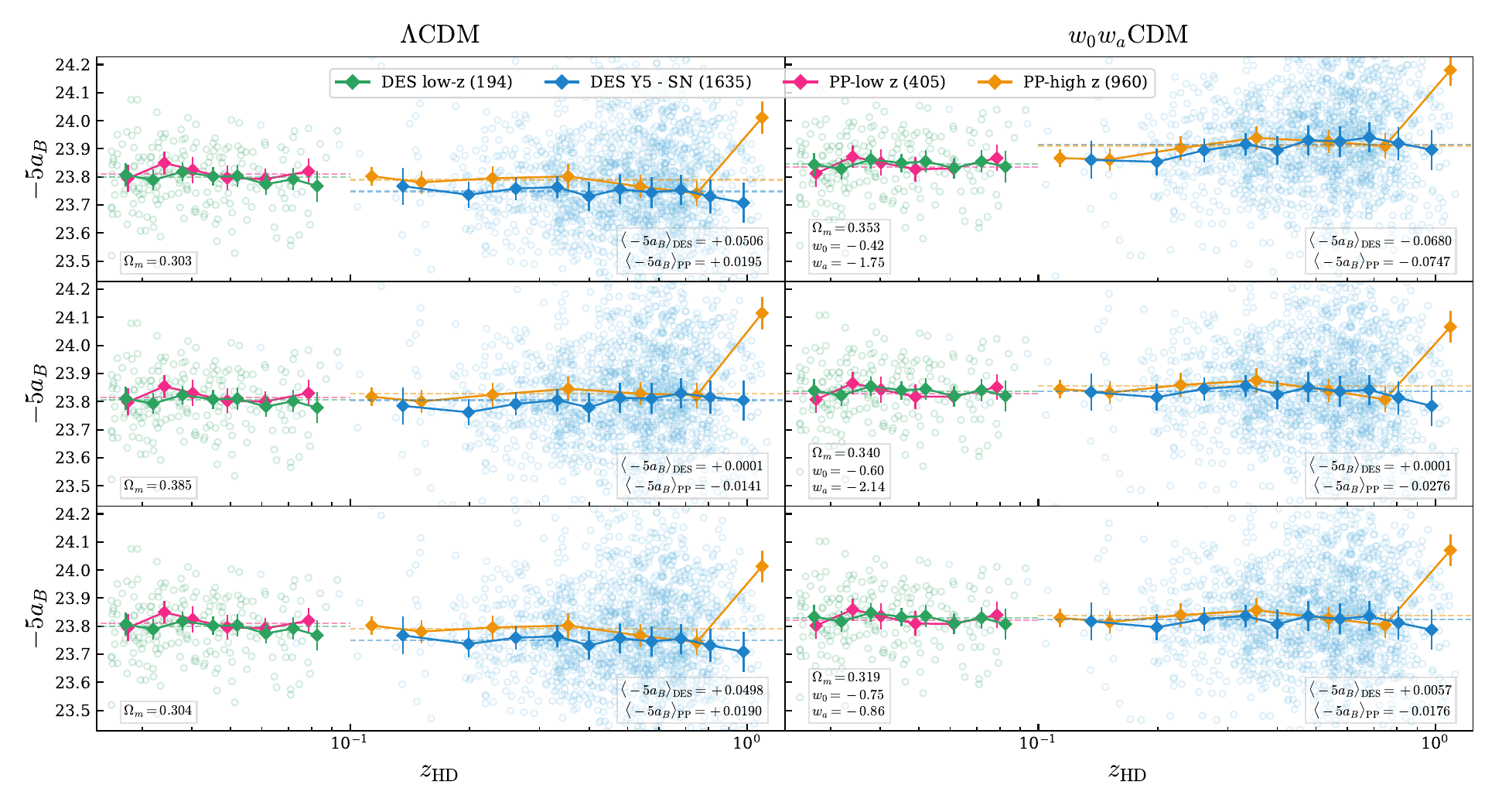}\\
    \includegraphics[width=1\linewidth]{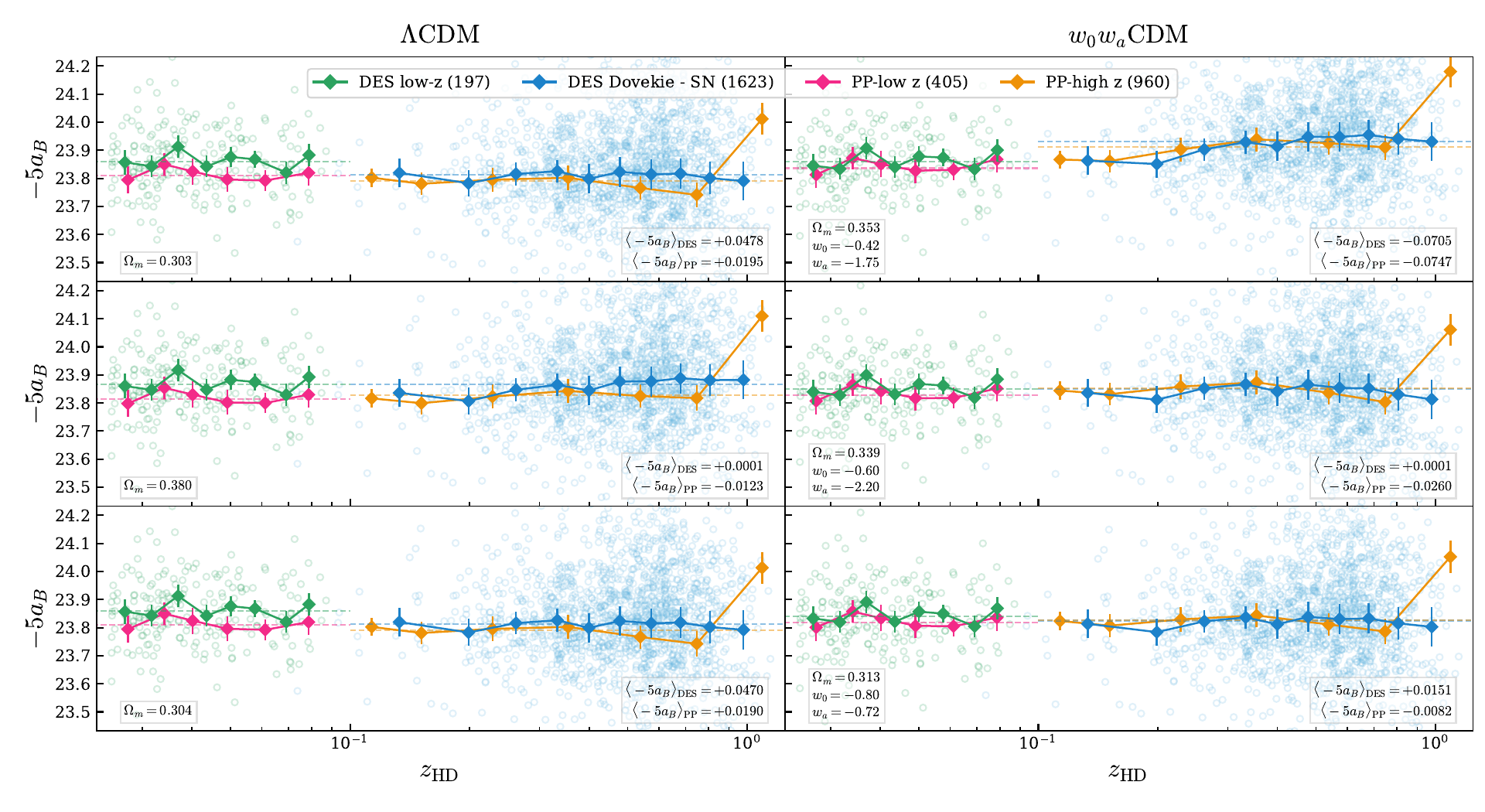}
    \caption{Re-analysis for the late-time $a_B$ tension with previous DESY5 (top) and updated DES-Dovekie (bottom)~\cite{Popovic:2025glk,DES:2025sig} compared to PantheonPlus, all of which are split into low-$z$ and high-$z$ parts around $z\sim0.1$. When computing $a_B$ from a specific $d_L(z)$, $\Lambda$CDM and $w_0w_a$CDM models are used in the left and right columns, respectively, where the parameter choices of corresponding models are as follows for both panels: the first row adopts the best-fit Planck-DESI constraints; the second row adopts those parameters within the $1\sigma$ range of best-fit Planck-DESI constraints to vanish the intercept discrepancy; the third row adopts the best-fit Planck-DESI-SNe constraints of corresponding SNe samples.}
    \label{fig:3.2}
\end{figure}

The recent DES 5YR update~\cite{Popovic:2025glk,DES:2025sig} effectively turns the original DES-SN5YR analysis into a fully recalibrated ``DES-Dovekie'' reanalysis: it improves the photometric cross-calibration using PS1/Gaia data and new DA white-dwarf observations, retrains the SALT3 light-curve model with the Dovekie calibration, corrects an outdated numerical approximation in the Fitzpatrick colour-law implementation, fixes an underweighted calibration-systematics contribution, regenerates simulations and bias corrections consistently with the updated SALT model, and reruns the full pipeline from light-curve fitting and photometric classification to covariance construction and cosmological inference. The calibration work shows that even small photometric shifts can propagate nontrivially into SN distances through colour–luminosity standardisation and SALT training, motivating the full end-to-end reanalysis rather than a simple parameter-level correction. 

Cosmologically, the update lowers the DES-SN5YR-inferred matter density in flat $\Lambda$CDM by $\Delta\Omega_m\simeq -0.022$ and, when combined with Planck+ACT+SPT CMB and DESI DR2 BAO in flat $w_0w_a$CDM, gives $w_0=-0.803\pm0.054$ and $w_a=-0.72\pm0.21$; the frequentist goodness-of-fit rejection significance of $\Lambda$CDM is reduced from the previous $\sim4.2\sigma$ DES-SN5YR-based result to $\sim3.2\sigma$, corresponding to only about $5:1$ weak Bayesian odds for evolving dark energy. Thus the updated DES 5YR analysis does not eliminate the preference for dynamical dark energy, but downgrades it from a strong-looking DESI-era anomaly to a weak-to-moderate indication that is significantly more sensitive to supernova calibration and modelling systematics. 

Note that recent re-analysis of Bayesian inference~\cite{Ong:2025utx,Ong:2026tta} using a public nested sampling code $\texttt{unimpeded}$~\cite{Ong:2025ver,Ong:2025cwv} have 
vanished the Bayesian evidence for dynamical dark energy, as the previous prior volume uses a somewhat more restricted and more phenomenology-shaped prior cube.
Here, we re-analyze our previous late-time $a_B$ tension with updated DES-Dovekie, as shown in the second panel of Fig.~\ref{fig:3.2}, compared to the first panel with DES-Y5. $\Lambda$CDM and $w_0w_a$CDM models are used in the left and right columns, respectively, with parameter choices as indicated in the figure caption.

For the first row of each panel with best-fit Planck-DESI parameters, it is easy to see that there is no intercept tension for PantheonPlus in the $\Lambda$CDM model, while for the $w_0w_a$CDM model, the $a_B$ tension appears for PantheonPlus,
\begin{align}
\langle-5a_B\rangle_\mathrm{PP}=+0.0195 \longrightarrow \langle-5a_B\rangle_\mathrm{PP}=-0.0747.
\end{align}
Furthermore, while there is an $a_B$ tension for DES-Y5 in the $\Lambda$CDM model, the same $a_B$ tension is worsen for DES-Y5 in the $w_0w_a$CDM model,
\begin{align}
\langle-5a_B\rangle_\mathrm{DES-Y5}=+0.0506 \longrightarrow \langle-5a_B\rangle_\mathrm{DES-Y5}=-0.0680.
\end{align}
Compared to DES-Y5, DES-Dovekie reduces the $a_B$ tension in the $\Lambda$CDM model, but it again worsens the $a_B$ tension in the $w_0w_a$CDM model,
\begin{align}
\langle-5a_B\rangle_\mathrm{DES-Dovekie}=+0.0478 \longrightarrow \langle-5a_B\rangle_\mathrm{DES-Dovekie}=-0.0705.
\end{align}
Only in the second row with parameters carefully chosen within the $1\sigma$ range of best-fit Planck-DESI constraints can we vanish the intercept discrepancy. Similar elimination of the $a_B$ tension in the third row of each panel is also achieved for the best-fit Planck-DESI-SNe constraints.

\begin{figure}
    \centering
    \includegraphics[width=1\linewidth]{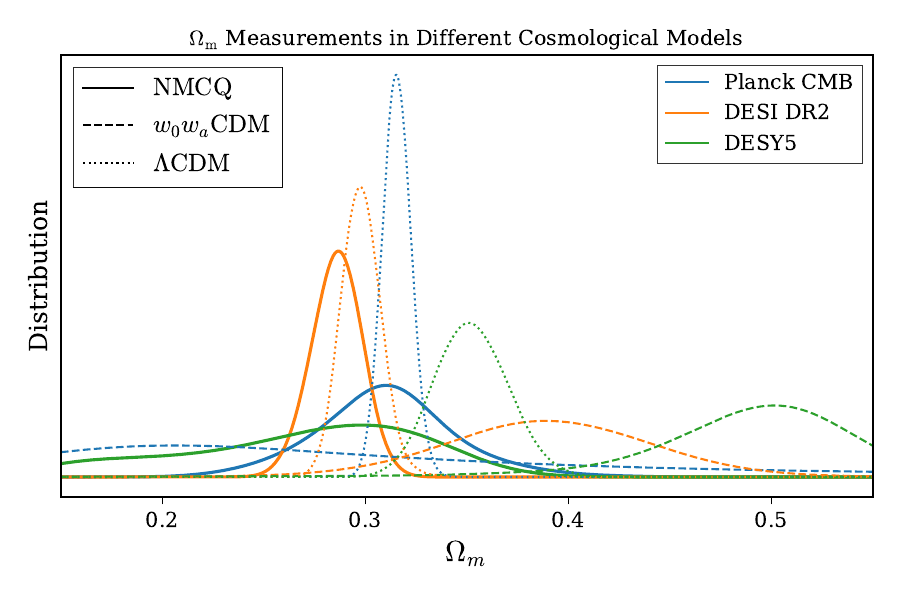}
    \caption{
    Comparison between dynamical and interacting dark-energy interpretations from Ref.~\cite{Wang:2025znm}. In the CPL $w_0w_a$CDM model, the separate Planck, DESI, and DES-Y5 constraints on $\Omega_{\rm m}$ remain widely dispersed, even though the combined fit improves over $\Lambda$CDM. In the nonminimally coupled quintessence model, the dark-sector interaction modifies the effective dark-matter evolution and brings the individual $\Omega_{\rm m}$ constraints into much better agreement.
    }
    \label{fig:4.1}
\end{figure}

This calls back a persistent tension in the supernovae intercept between DES-Y5/Dovekie and Planck-DESI datasets, and the preference for the $w_0w_a$CDM model emerges largely as a compromise of inter-data tension, for example, the matter fraction ($\Omega_m$) tension among Planck, DESI, and DESY5, as shown for both $\Lambda$CDM and $w_0w_a$CDM models in Fig.~\ref{fig:4.1}, where $w_0w_a$CDM model achieves a better fit to $\Omega_m$ than $\Lambda$CDM among Planck, DESI, and DESY5 only via reduced and enlarged distributions. In the next section, we will show that it is possible to re-concentrate the $\Omega_m$ distribution with interacting dark energy models.

\section{Late-time dark energy}\label{sec:DE}

\subsection{Dynamical vs. interacting dark energy}\label{subsec:DDEIDE}

The late-time $a_B$ analysis above suggests that part of the apparent preference for dynamical dark energy can be traced to supernova calibration and low-redshift anchoring systematics. Nevertheless, there remains a more general issue in the combined interpretation of Planck CMB, DESI BAO, and DES-Y5 SNe: in simple late-time parameterizations, the datasets do not prefer the same matter fraction $\Omega_{\rm m}$. In $\Lambda$CDM, Planck prefers a higher matter fraction than DESI BAO, while DES-Y5 prefers an even higher value than both. This mismatch becomes more severe when the dark-energy equation of state is promoted to the CPL form $w(a)=w_0+w_a(1-a)$. Although the $w_0w_a$CDM model can reduce the total $\chi^2$ and mimic the DESI-preferred phantom-crossing behavior, it does so partly by enlarging the allowed parameter volume rather than by genuinely aligning the individual dataset preferences. In this sense, the usual dynamical-dark-energy interpretation should be distinguished from a physical resolution of the underlying CMB--BAO--SN matter-fraction tension.

A useful alternative is to allow the dark matter sector to interact with a quintessence field. In the nonminimally coupled quintessence (NMCQ) model considered in Ref.~\cite{Wang:2025bkk}, standard-model particles remain minimally coupled to the Einstein-frame metric $g_{\mu\nu}$, while dark matter couples to the conformally related metric $\tilde g_{\mu\nu}={\cal A}^2(\varphi)g_{\mu\nu}$. The total action may be written schematically as
\begin{align}
    S=S_{\rm GR}
    +S_{\rm SM}[\psi_{\rm SM};g_{\mu\nu}]
    +S_{\rm DM}[\psi_{\rm DM};{\cal A}^2(\varphi)g_{\mu\nu}]
    +S_\varphi ,
\end{align}
with
\begin{align}
    S_\varphi=\int d^4x\sqrt{-g}
    \left[
        -\frac12 g^{\mu\nu}\partial_\mu\varphi\partial_\nu\varphi
        -V(\varphi)
    \right].
\end{align}
A representative realization uses the dilaton coupling and Ratra--Peebles potential
\begin{align}
    {\cal A}(\varphi)=e^{-\beta\varphi/M_{\rm Pl}},
    \qquad
    V(\varphi)=\alpha\Lambda^4
    \left(\frac{\varphi}{M_{\rm Pl}}\right)^{-n}.
\end{align}
The limit $\beta=n=0$ recovers $\Lambda$CDM. Since the scalar-mediated force acts only in the dark sector, the model is not subject to ordinary baryonic fifth-force bounds.

The background equations are
\begin{align}
    \rho_{\rm r}+\rho_{\rm b}+\rho_{\rm DM}+\rho_\varphi
    &=3M_{\rm Pl}^2H^2,\\
    \dot\rho_\varphi+3H(1+w_\varphi)\rho_\varphi
    &=-\frac{{\cal A}'(\varphi)}{{\cal A}(\varphi)}
    \dot\varphi\,\rho_{\rm DM},\\
    \dot\rho_{\rm DM}+3H\rho_{\rm DM}
    &=+\frac{{\cal A}'(\varphi)}{{\cal A}(\varphi)}
    \dot\varphi\,\rho_{\rm DM}.
\end{align}
Thus, the Einstein-frame dark matter density no longer scales exactly as $a^{-3}$, but instead obeys
\begin{align}
    \frac{\rho_{\rm DM}}{\rho_{{\rm DM},0}}=
    \left(\frac{a}{a_0}\right)^{-3}
    \frac{{\cal A}(\varphi)}{{\cal A}(\varphi_0)} .
    \label{eq:NMCQ-rhoDM}
\end{align}
This modification is the essential difference from a purely phenomenological dynamical-dark-energy fit: the dark-sector interaction changes the effective matter evolution itself, and hence can directly address the $\Omega_{\rm m}$ mismatch among CMB, BAO, and SN data, and also affect our interpretation of DESI dynamical dark energy, as we will see in the next subsection.

For the combined Planck+DESI+DES-Y5 analysis, the CPL model gives $w_0=-0.756\pm0.057$ and $w_a=-0.840^{+0.220}_{-0.225}$ with $\Omega_{\rm m}=0.319\pm0.0055$ and $S_8=0.827\pm0.009$. It improves the best-fit value relative to $\Lambda$CDM by $\Delta\chi^2=-17.9$ with a Bayes factor $\ln{\cal B}=+3.69\pm0.30$. However, when Planck, DESI, and DES-Y5 are fitted separately, their preferred $\Omega_{\rm m}$ values remain widely dispersed in the CPL parameterization. In particular, the separate best-fit values move roughly toward
\begin{align}
    \Omega_{\rm m}^{\rm Planck}\simeq0.150,
    \qquad
    \Omega_{\rm m}^{\rm DESI}\simeq0.390,
    \qquad
    \Omega_{\rm m}^{\rm DESY5}\simeq0.504 ,
\end{align}
showing that the apparent success of $w_0w_a$CDM partly comes from weakened constraints and enlarged degeneracy directions.

By contrast, the nonminimally coupled quintessence model gives $n=0.62\pm0.18$ and $\beta=0.054^{+0.012}_{-0.008}$ with more than $3\sigma$ evidence for a nonzero coupling. Note that the very same model~\cite{Wang:2025bkk}, when further fitted to ACT and SPT, would raise the significance of a nonzero coupling up to $5\sigma$~\cite{Li:2026xaz}. The combined constraint is $\Omega_{\rm m}=0.302\pm0.0053$ and $S_8=0.820\pm0.008$, and the model improves over $\Lambda$CDM by $\Delta\chi^2=-12.4$ with a Bayes factor $\ln{\cal B}=+2.66\pm0.30$. Although the Bayesian preference is slightly weaker than that of the CPL fit, the physical behavior is cleaner: the separate Planck, DESI, and DES-Y5 constraints on $\Omega_{\rm m}$ become much more mutually consistent. Note that in a more recent construction of interacting dark energy~\cite{Wang:2026wrk}, even the Bayes evidence for every data combination performs better than the $w_0w_a$CDM model. Their best-fit values are approximately
\begin{align}
    \Omega_{\rm m}^{\rm Planck}\simeq0.311,
    \qquad
    \Omega_{\rm m}^{\rm DESI}\simeq0.286,
    \qquad
    \Omega_{\rm m}^{\rm DESY5}\simeq0.304,
\end{align}
so that the combined value lies within the allowed regions of the individual datasets. This is the main sense in which an interacting dark-sector model can resolve, rather than merely dilute, the Planck--DESI--DES-Y5 matter-fraction tension.

The physical lesson is that the DESI preference for phantom crossing need not be interpreted literally as a single minimally coupled dark-energy fluid crossing $w=-1$. Such a crossing is difficult to realize in a stable single-fluid model (see, however, the quintom model~\cite{Feng:2004ad,Feng:2004ff,Guo:2004fq}). Instead, it may be an apparent behavior produced when a more complicated dark sector, containing interacting dark matter and quintessence, is projected onto the simpler CPL template. In the interacting description, the energy transfer changes the effective matter evolution, while in the CPL description, the same effect is absorbed into an artificial time-dependent dark-energy equation of state.

\subsection{Effective vs. apparent dark energy}\label{subsec:effappDE}

\begin{figure}
    \centering
    \includegraphics[width=1\linewidth]{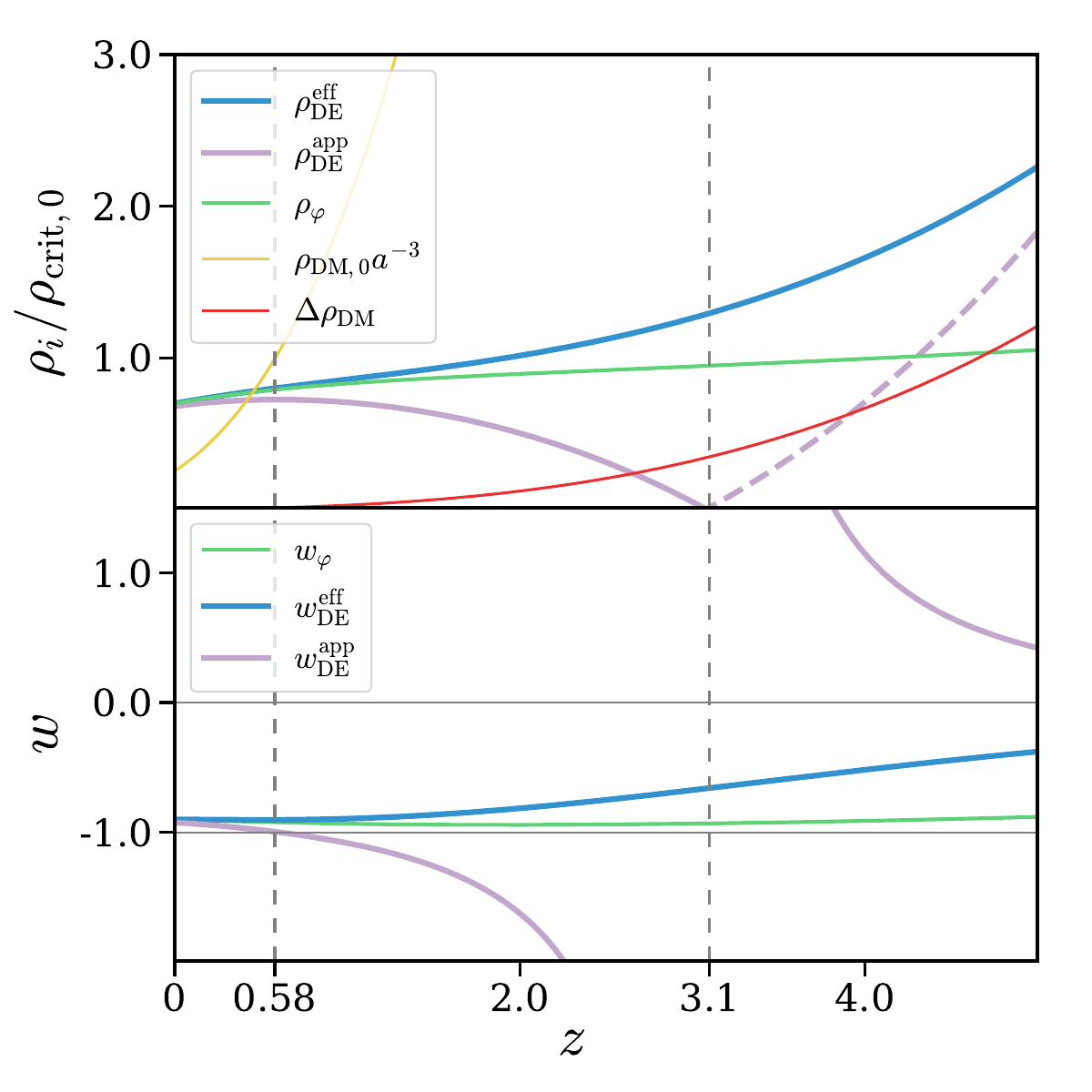}
    \caption{
    Full redshift evolution of the physical and reconstructed dark-sector components from Ref.~\cite{Wang:2025znm}. The effective dark energy $\rho_{\rm DE}^{\rm eff}$ remains nonphantom, while the apparent dark energy $\rho_{\rm DE}^{\rm app}$ inferred by matching the NMCQ model to a CPL decomposition can cross $w=-1$ around $z\simeq0.5$. At higher redshift, the apparent density may become negative and its equation of state may diverge, but this is only a decomposition artifact caused by assigning the mismatched dark-matter density to the dark-energy sector.
    }
    \label{fig:4.2}
\end{figure}

The comparison above shows that the nonminimally coupled quintessence model can align the matter fractions preferred by Planck, DESI, and DES-Y5 more efficiently than the phenomenological CPL model. There is, however, an apparent puzzle to address. The DESI-motivated CPL fit prefers a phantom-crossing behavior, while the underlying scalar field in the interacting model is an ordinary quintessence field with $w_\varphi>-1$. The resolution is that the equation of state inferred by a CPL fit is not necessarily the physical equation of state of the scalar field, nor even the usual effective equation of state of interacting dark energy. It is an apparent equation of state obtained after forcing the true interacting dark sector into a mismatched noninteracting CDM plus CPL-dark-energy decomposition. We explain this long-neglected subtlety below.

The dark-sector interaction gives
\begin{align}
    \dot{\rho}_\varphi+3H(1+w_\varphi)\rho_\varphi&=-Q,\\
    \dot{\rho}_{\rm DM}+3H\rho_{\rm DM}&=+Q,
\end{align}
where
\begin{align}
    Q=\frac{{\cal A}'(\varphi)}{{\cal A}(\varphi)}\dot\varphi\,\rho_{\rm DM}.
\end{align}
Since the interacting dark matter density does not scale exactly as $a^{-3}$, one may artificially subtract a would-be standard CDM component,
\begin{align}
    \rho_{\rm DM}^{\rm fid}=\rho_{\rm DM,fid}a^{-3},
\end{align}
and define the residual noncold dark-matter contribution as
\begin{align}
    \Delta\rho_{\rm DM}\equiv\rho_{\rm DM}-\rho_{\rm DM}^{\rm fid}.
\end{align}
The usual effective dark-energy density is then
\begin{align}
    \rho_{\rm DE}^{\rm eff}=\rho_\varphi+\Delta\rho_{\rm DM},
    \label{eq:rhoDEeff}
\end{align}
with
\begin{align}
    \dot{\rho}_{\rm DE}^{\rm eff}+3H(1+w_{\rm DE}^{\rm eff})\rho_{\rm DE}^{\rm eff}=0 .
\end{align}
Therefore, the usual effective equation of state reads
\begin{align}
    w_{\rm DE}^{\rm eff}
    =\frac{w_\varphi\rho_\varphi}{\rho_{\rm DE}^{\rm eff}}
    =\frac{w_\varphi}{1+(\rho_{\rm DM}-\rho_{\rm DM}^{\rm fid})/\rho_\varphi}.
    \label{eq:wDEeff}
\end{align}
For the dilaton coupling used here, ${\cal A}(\varphi)=\exp(-\beta\varphi/M_{\rm Pl})$, the residual contribution can remain positive for the best-fit evolution when the fiducial component is chosen as $\rho_{\rm DM}^{\rm fid}=\rho_{\rm DM,0}a^{-3}$ (iff you expect all dark matter today is cold, that is, there is no noncold dark matter at present day, $\Delta\rho_{\rm DM,0}=\rho_{\rm DM,0}-\rho_{\rm DM,0}^{\rm fid}=0$). Then the effective equation of state stays above the quintessence equation of state and never crosses the phantom divide, $w_{\rm DE}^{\rm eff}\geq w_\varphi>-1$. Thus, the physical interacting model itself does not contain a phantom scalar or a real phantom instability.

The apparent equation of state relevant for interpreting a CPL fit is different. If the same total dark-sector density is instead decomposed as a CPL dark-energy component plus the CDM component preferred by the $w_0w_a$CDM fit, then
\begin{align}
    \rho_\varphi+\rho_{\rm DM}=\rho_{\rm DE}^{\rm app}+\rho_{\rm CDM}^{\rm CPL},
\end{align}
with
\begin{align}
    \rho_{\rm CDM}^{\rm CPL}=\rho_{\rm CDM,0}^{\rm CPL}a^{-3}.
\end{align}
This defines the apparent dark-energy density
\begin{align}
    \rho_{\rm DE}^{\rm app}=
    \rho_\varphi+\rho_{\rm DM}-
    \rho_{\rm CDM}^{\rm CPL}=
    \rho_{\rm DE}^{\rm eff}+
    \rho_{\rm DM}^{\rm fid}-
    \rho_{\rm CDM}^{\rm CPL}.
    \label{eq:rhoDEapp}
\end{align}
Its equation of state is
\begin{align}
    w_{\rm DE}^{\rm app}
    =\frac{w_{\rm DE}^{\rm eff}\rho_{\rm DE}^{\rm eff}}{\rho_{\rm DE}^{\rm app}}=\frac{w_\varphi\rho_\varphi}{\rho_{\rm DE}^{\rm app}}
    =\frac{w_\varphi}{1+(\rho_{\rm DM}-\rho_{\rm CDM}^{\rm CPL})/\rho_\varphi} .
    \label{eq:wDEapp}
\end{align}
The relation $w_\varphi\rho_\varphi=w_{\rm DE}^{\rm eff}\rho_{\rm DE}^{\rm eff}=w_{\rm DE}^{\rm app}\rho_{\rm DE}^{\rm app}$ simply states that the subtracted would-be standard-CDM pieces carry no pressure. The crucial point is that $w_{\rm DE}^{\rm app}$ depends on the CDM density assumed by the comparison model, not only on the physical scalar field. Therefore, if the CPL fit assigns a different matter fraction from the NMCQ model, the mismatch
\begin{align}
    \rho_{\rm DM}^{\rm NMCQ}-\rho_{\rm CDM}^{\rm CPL}
\end{align}
is reassigned from the dark-matter sector to the dark-energy sector. This reassignment can make the apparent density evolve as if it crossed the phantom divide, even when the effective density in Eq.~\eqref{eq:rhoDEeff} remains nonphantom. This is the key point of this proceeding review.

This explains the full redshift evolution shown in Fig.~\ref{fig:4.2}. At late times, $0<z\lesssim0.45$, the quintessence component dominates the dark sector and behaves approximately like a cosmological constant. The apparent equation of state is then close to $-1$ and may lie above $-1$. Around the transition from dark-energy domination to dark-matter domination, the mismatch between $\rho_{\rm DM}^{\rm NMCQ}$ and $\rho_{\rm CDM}^{\rm CPL}$ becomes important. The apparent density $\rho_{\rm DE}^{\rm app}$ changes its time-derivative behavior, so the apparent equation of state crosses $w_{\rm DE}^{\rm app}=-1$ around $z\simeq0.5$, reproducing the phantom-crossing behavior found when the data are interpreted with the CPL parameterization.

At higher redshifts, the matter component dominates. The mismatch term can become larger than the scalar-field energy density, and the apparent dark-energy density can even pass through zero around $z\simeq3.1$. The corresponding divergence of $w_{\rm DE}^{\rm app}$ is not a physical singularity. It only reflects the fact that the artificial component $\rho_{\rm DE}^{\rm app}$ is defined by subtracting the wrong CDM density from the true interacting dark sector. The physical quantities actually evolved in the model, namely $\rho_\varphi$ and $\rho_{\rm DM}$, remain regular.

Finally, at very high redshifts, the residual noncold dark-matter contribution tracks the difference needed to recover the Planck-$\Lambda$CDM matter density near recombination. In this way, the NMCQ model can preserve the CMB-era matter budget while modifying the late-time mapping between BAO, SNe, and the inferred matter fraction. The phantom crossing is therefore not a fundamental property of the dark-energy field; it is an apparent behavior produced when an interacting dark sector is projected onto a noninteracting CPL template.

\section{Conclusions and discussions}\label{sec:condis}

In this mini-review, we have summarized the recent development of the cosmological intercept tension from the perspective of the standardized supernova magnitude--redshift relation. The central conclusion is that the intercept is not merely a nuisance parameter describing the degeneracy between the Hubble constant and the standardized absolute magnitude of Type Ia supernovae, but a directly reconstructable consistency diagnostic of the supernova Hubble diagram. 

Applied to PantheonPlus, this diagnostic reveals a robust local intercept tension that persists across different late-time cosmological parameterizations and calibration strategies. A more detailed decomposition shows that the dominant displacement is mainly associated with the Cepheid-hosted second-rung supernovae rather than the full local sample, while non-Cepheid local supernovae remain broadly consistent with the third-rung Hubble-flow intercept. Imposing intercept consistency as a redshift-correction criterion restores agreement between the first two rungs and the SH0ES-like local distance-ladder determination, thereby sharpening rather than removing the Hubble tension: the remaining discrepancy is between the early-Universe CMB inference and the first two rungs of the local distance ladder, not a failure of the third-rung Hubble-flow supernovae. 

The same intercept logic also clarifies the DESI-era dark-energy anomaly. The DES-Y5 external low-redshift supernova subset shows an intercept offset and irregular low-redshift fluctuations relative to the DES-SN subset, a feature not seen as strongly in PantheonPlus. Restoring low-redshift and DES-SN intercept consistency substantially reduces the apparent preference for CPL-like dynamical dark energy, and the updated DES-Dovekie reanalysis further downgrades this preference by improving the photometric calibration, light-curve-model training, colour-law treatment, calibration covariance, and simulation-based bias corrections. Thus, the current DESI plus supernova evidence for evolving dark energy should be interpreted cautiously, since part of it is sensitive to low-redshift supernova anchoring and calibration systematics.

The broader discussion is that intercept consistency provides a useful organizing principle for deciding when cosmological tensions require new physics and when they instead point to residual data-systematics issues. In the Hubble-tension context, it separates the local distance-ladder problem into second-rung calibration, third-rung Hubble-flow consistency, and early-Universe CMB inference. In the DESI dark-energy context, it separates a possible homogeneous late-time transition from low-redshift supernova-anchor systematics. This distinction is especially important because phenomenological dynamical-dark-energy parameterizations can improve the global fit partly by enlarging degeneracy directions and compromising among datasets that prefer different matter fractions, rather than by genuinely reconciling the individual dataset preferences. 

Interacting dark energy offers a physically motivated alternative interpretation: by modifying the effective dark-matter evolution, it can align the matter fractions inferred from CMB, BAO, and supernova data more directly, while the apparent phantom crossing seen in a noninteracting CPL fit can emerge as a projection effect rather than a fundamental phantom component. Future progress, therefore, requires end-to-end supernova consistency tests using common calibration, covariance construction, redshift corrections, selection modelling, and light-curve standardization across PantheonPlus, DES-Y5/DES-Dovekie, and future low-redshift anchor samples. On the theory side, phenomenological fits should be compared with physical dark-sector models according to whether they reconcile the separate datasets, not merely whether they improve the combined goodness of fit.

\begin{acknowledgments}
We are grateful for the early collaborations with Dr. Lu Huang on this project, who has left the scientific community to pursue a better life.
This work is supported by the National Key Research and Development Program of China Grants No. 2021YFC2203004 and No. 2021YFA0718304, the National Natural Science Foundation of China Grants No. 12422502, No. 12547110, No.12588101, No. 12235019, and No. 12447101, and the China Manned Space Program Grant No. CMS-CSST-2025-A01.
\end{acknowledgments}

\bibliographystyle{JHEP}
\bibliography{refer}

\end{document}